\documentclass{article}

\usepackage[preprint]{neurips_2025}

\usepackage[utf8]{inputenc} 
\usepackage[T1]{fontenc}    
\usepackage{hyperref}       
\usepackage{url}            
\usepackage{booktabs}       
\usepackage{amsfonts}       
\usepackage{nicefrac}       
\usepackage{microtype}      
\usepackage{xcolor}         
\usepackage{graphicx}
\usepackage{amsmath}
\usepackage{multirow}
\usepackage{bbding}
\usepackage{bm}
\usepackage{braket}
\usepackage{listings}
\usepackage{textcomp}
\usepackage{makecell}

\graphicspath{{./figs/}}

\title{MIRB: Mathematical Information Retrieval Benchmark}

\author{
  \textbf{Haocheng Ju\textsuperscript{1}},
  \textbf{Bin Dong,\textsuperscript{2,3,4}}\thanks{Corresponding author}
\\
\\
 \textsuperscript{1}School of Mathematical Sciences, Peking University\\
 \textsuperscript{2}Beijing International Center for Mathematical Research\\ and the New Cornerstone Science Laboratory, Peking University\\
  \textsuperscript{3}Center for Machine Learning Research, Peking University\\
  \textsuperscript{4}Center for Intelligent Computing, Great Bay Institute for Advanced Study, \\Great Bay University
\\
 \texttt{hcju@pku.edu.cn} \qquad
 \texttt{dongbin@math.pku.edu.cn}
}

\begin{document}

\maketitle

\begin{abstract}
  Mathematical Information Retrieval (MIR) is the task of retrieving information from mathematical documents and plays a key role in various applications, including theorem search in mathematical libraries, answer retrieval on math forums, and premise selection in automated theorem proving. However, a unified benchmark for evaluating these diverse retrieval tasks has been lacking. In this paper, we introduce MIRB (Mathematical Information Retrieval Benchmark) to assess the MIR capabilities of retrieval models. MIRB includes four tasks—semantic statement retrieval, question-answer retrieval, premise retrieval, and formula retrieval—spanning a total of 12 datasets. We evaluate 13 retrieval models on this benchmark and analyze the challenges inherent to MIR. We hope that MIRB provides a comprehensive framework for evaluating MIR systems and helps advance the development of more effective retrieval models tailored to the mathematical domain.\footnote{Our code and data are available at \href{https://github.com/j991222/mirb}{https://github.com/j991222/mirb} and \href{https://huggingface.co/collections/hcju/mirb-6827001711765454f58c5a76}{https://huggingface.co/collections/hcju/mirb-6827001711765454f58c5a76}}
\end{abstract}

\section{Introduction}
Mathematical Information Retrieval (MIR) \cite{dadure2024mathematical,zanibbi2025mathematical} focuses on retrieving mathematical content such as definitions, theorems, and proofs from a mathematical corpus. MIR has many practical applications. For instance, mathematicians working with Lean \citep{DBLP:conf/cade/MouraKADR15,DBLP:conf/cade/Moura021} often need to verify whether a particular theorem exists in mathlib4, Lean’s mathematical library. In this case, the MIR query can be either a natural language or formal statement, and the corpus consists of declarations in mathlib4. Another example is students searching for similar questions or answers on Mathematics Stack Exchange to help them solve problems. Here, the user’s question serves as the query, and the corpus includes all question and answer posts on the forum. MIR is also an essential component in automated theorem proving, in both natural and formal languages. For example, NaturalProver \cite{welleck2022naturalprover} is a natural language theorem prover that uses stepwise beam search to sample proofs, retrieving multiple references from a corpus of ProofWiki definitions and theorems to support reliable tactic generation. Similarly, ReProver \cite{yang2023leandojo} is a formal theorem prover for Lean that performs best-first search; at each step, it retrieves premises from mathlib4 using the current proof state as the query, and feeds the retrieved premises into a tactic generator. This retrieval step is often referred to as premise retrieval. In summary, MIR plays a crucial role in a wide range of mathematical applications.

MIR differs from standard text retrieval in that both queries and documents often contain mathematical formulas. These formulas are highly structured, and their semantic meaning typically remains unchanged under variable substitution, even though their textual representations differ. This structural property poses unique challenges for retrieval models, which must adapt to the specific characteristics of mathematical language. Due to the importance of MIR, several competitions have been organized to evaluate different MIR systems. For example, ARQMath \cite{10.1007/978-3-030-58219-7_15,10.1007/978-3-030-85251-1_17,10.1007/978-3-030-99739-7_51}, held at the Conference and Labs of the Evaluation Forum (CLEF) from 2020 to 2022, includes two main tasks: answer retrieval and formula retrieval, with both queries and corpora sourced from Mathematics Stack Exchange. Similarly, the NTCIR series \cite{zanibbi2016ntcir} features a formula+keyword search task over corpora drawn from arXiv and Wikipedia. However, existing MIR datasets are limited in both task diversity and domain coverage, and are scattered across different sources. To the best of our knowledge, there is no unified benchmark that consolidates all major MIR tasks and datasets for a comprehensive evaluation of retrieval models.

To address this gap, we introduce \textbf{MIRB} (Mathematical Information Retrieval Benchmark), a comprehensive benchmark designed to assess retrieval models on a wide range of MIR tasks across various domains and languages. MIRB covers four main tasks: Semantic Statement Retrieval, Question Answer Retrieval, Premise Retrieval, and Formula Retrieval, across 12 datasets in diverse mathematical domains and languages. We evaluate 13 retrieval models on this benchmark and observe that all models perform worse on reasoning-based tasks compared to semantic-based tasks. Moreover, applying cross-encoder rerankers generally leads to performance degradation. These results highlight that current retrieval models still have much room for improvement in handling MIR tasks.

The rest of the paper is organized as follows. We review the related works on retrieval benchmarks, retrieval models and mathematical information retrieval in Section \ref{sec:related}. Section \ref{sec:mirb} describe the tasks included in MIRB and the details of the dataset construction process. Experimental results of the evaluated retrieval models are presented in Section \ref{sec:exp}, and the paper concludes in Section \ref{sec:conclusion}.

\section{Related Work}\label{sec:related}
\subsection{Retrieval Benchmarks}
Existing retrieval benchmarks can generally be divided into two categories: (1) general-purpose benchmarks that span diverse domains and tasks, such as BEIR \cite{thakur2021beir}, MTEB \cite{muennighoff2022mteb}, MMTEB \cite{enevoldsen2025mmteb}, C-MTEB \cite{xiao2024c} and MAIR \cite{sun2024mair}; and (2) domain-specific or task-specific benchmarks that focus on a particular domain or retrieval task. For example, ChemTEB \cite{kasmaee2024chemteb} includes a retrieval benchmark for chemistry, while CodeSearchNet \cite{husain2019codesearchnet}, CosQA \cite{huang2021cosqa}, XcodeEval \cite{khan2023xcodeeval}, and CoIR \cite{li2024coir} target code retrieval. LONGEMBED \cite{zhu2024longembed} is designed for long-context retrieval. The benchmarks most closely related to our work are RAR-b \cite{xiao2024rar} and BRIGHT \cite{su2025bright}, both of which include reasoning-based retrieval datasets covering commonsense reasoning, mathematics, and code. In RAR-b's question-answer retrieval task, relevant documents directly answer the query, while BRIGHT focuses on retrieving documents that either assist in answering the query or use the same theorem as the one in the query. Our work differs from these benchmarks in three aspects: (1) we focus exclusively on the mathematics domain; (2) we include both semantic retrieval tasks (Semantic Statement Retrieval, Formula Retrieval) and reasoning-based tasks (Question-Answer Retrieval, Premise Retrieval), whereas RAR-b and BRIGHT focus solely on reasoning-based retrieval; (3) within reasoning-based retrieval, we include the task of premise retrieval in both natural and formal language, which is not covered in either RAR-b or BRIGHT.

\subsection{Retrieval Models}
The development of retrieval models has advanced beyond the classic BM25 algorithm \cite{robertson1995okapi,10.1561/1500000019}, which relies on sparse vector representations and measures lexical similarity between queries and documents. Modern approaches leverage deep neural networks to encode queries and documents into dense vectors, enabling relevance assessment based on semantic similarity. A widely adopted training paradigm for these dense retrieval models \cite{DBLP:journals/corr/abs-2201-10005,DBLP:journals/corr/abs-2212-03533,su-etal-2023-one,xiao2024c} involves pretraining on large-scale unsupervised data using contrastive loss, followed by fine-tuning on smaller labeled datasets. In terms of architecture, earlier models commonly employed bidirectional encoders, but recent studies \cite{wang-etal-2024-improving-text,SFR-embedding-2,SFRAIResearch2024,lee2025nvembed} have demonstrated that decoder-only language models can achieve superior performance. Moreover, the training data for retrieval models can be augmented with synthetic data generated by large language models \cite{wang-etal-2024-improving-text,muennighoff2024generative,lee2024gecko}.

\subsection{Mathematical Information Retrieval.}
Classical mathematical information retrieval methods often rely on tree-based representations to capture the structural information of mathematical formulas, such as the Symbol Layout Tree\cite{DBLP:journals/ijdar/ZanibbiB12} and the Operator Tree \cite{DBLP:conf/ntcir/GaoYWJT16}. A representative approach is the structure search used in Approach0 \cite{DBLP:conf/ecir/ZhongZ19,10.1007/978-3-030-45439-5_47}, which computes structural similarity by identifying the largest common subexpressions and matching maximum subtrees. More recent methods combine structure-based search with dense retrieval models \cite{DBLP:conf/clef/KaneNT22,DBLP:conf/clef/Zhong0L22,DBLP:conf/emnlp/ZhongY0L22,10.1145/3539618.3591746}, allowing systems to handle both the semantic similarity of text and the structural similarity of formulas. In general, dense retrievers such as text embedding models are more robust to invalid LaTeX formulas and to formulas written in alternative formats, whereas traditional structure based methods often fail at the parsing stage if the LaTeX syntax is incorrect.

\begin{figure}[tb]
\centering
\includegraphics[width=0.9\linewidth]{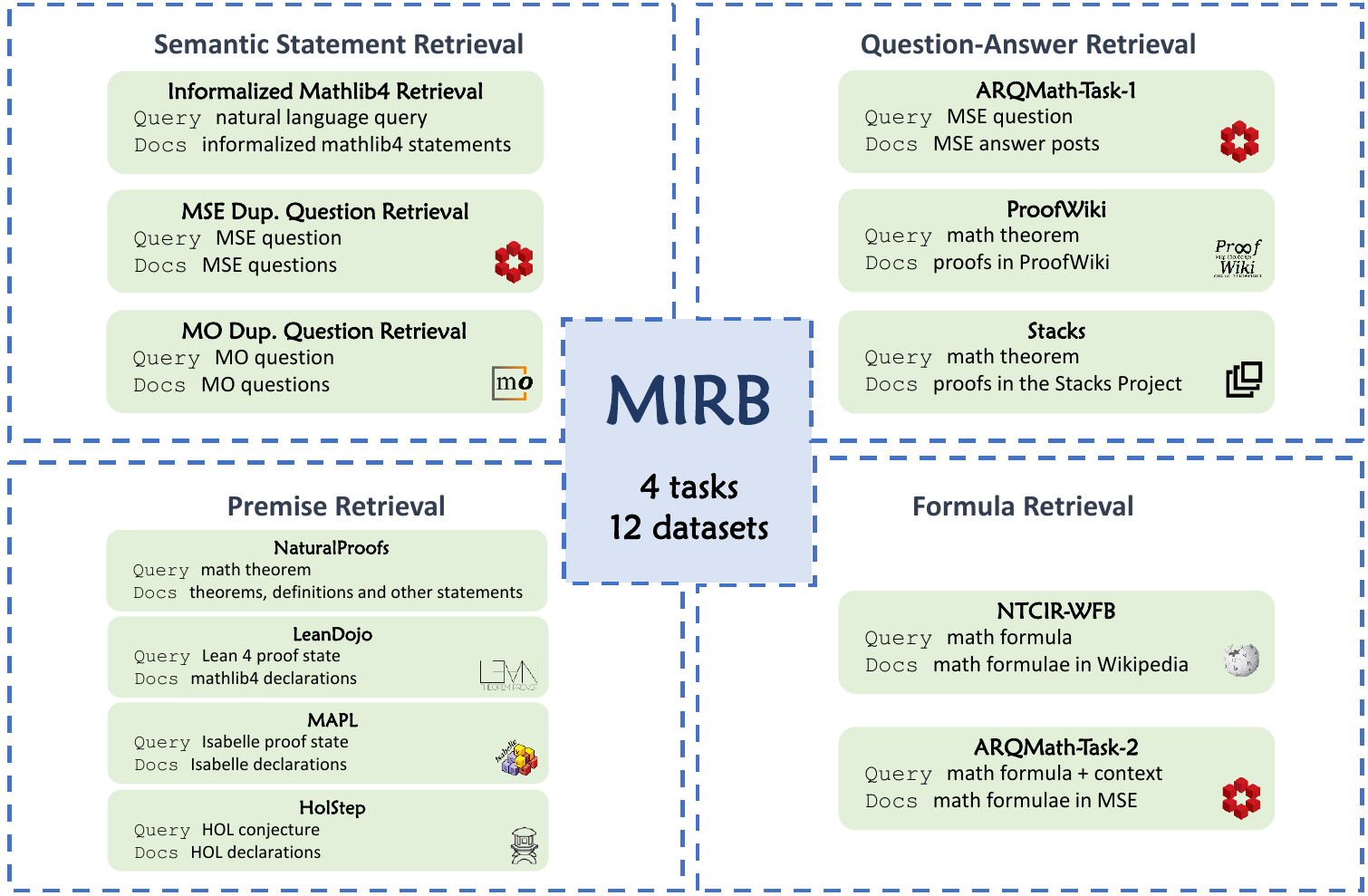}
\caption{Overview of tasks and datasets in MIRB.}
\label{fig:overview}
\end{figure}

\section{The MIRB Benchmark}\label{sec:mirb}
We present \textbf{MIRB}, a benchmark designed to evaluate the mathematical information retrieval capabilities of retrieval models. It comprises four tasks: Semantic Statement Retrieval, Question-Answer Retrieval, Premise Retrieval and Formula Retrieval. Dataset statistics are provided in Table \ref{tab:datasets_info}. The following four subsections describe each task and the corresponding dataset construction in detail.

\begin{table}[tb]
\caption{Statistics of the datasets. We report the number of queries and documents in each dataset. \textbf{Avg. D / Q} denotes the average number of relevant documents per query. Average Word Length refers to the mean number of words per query or per document. Examples from five representative datasets (Informalized Mathlib4 Retrieval, MSE Dup. Question Retrieval, ARQMath-Task-1, NaturalProofs, NTCIR-WFB) are included in the main text, while examples from the remaining datasets are provided in the appendix.}
\label{tab:datasets_info}
\centering
\resizebox{\linewidth}{!}{
\begin{tabular}{llccccccc}
\toprule
& &   & \multicolumn{3}{c}{\textbf{Test}}   & \multicolumn{2}{c}{\textbf{Avg. Word Length}}       \\
\cmidrule(lr){4-6}\cmidrule(l){7-8}
\textbf{Task} & \textbf{Dataset} & \textbf{Relevancy} & \textbf{\#query} & \textbf{\#corpus}  & \textbf{Avg. D / Q}  & \textbf{Query} & \textbf{Document} & \textbf{Example} \\
\midrule
\multirow{2}{*}{Semantic Statement Retrieval} 
& Informalized Mathlib4 Retrieval \cite{gao-etal-2024-semantic-search} & 3-level & 40 & 124,254 & 7.23 & 10.38 &  41.60 & Table \ref{tab:inf_mathlib} \\
& MSE Dup. Question Retrieval & Binary & 25,116 & 1,350,505 & 1.78  & 97.22 & 116.42 & Table \ref{tab:msedup}\\
& MO Dup. Question Retrieval        & Binary  & 225   & 108,301 & 1.08 & 100.78 & 144.53 & Table \ref{tab:modup} \\
\midrule
\multirow{3}{*}{Question-Answer Retrieval} 
& ARQMath-Task-1 \cite{10.1007/978-3-030-58219-7_15,10.1007/978-3-030-85251-1_17,10.1007/978-3-030-99739-7_51} & 4-level & 78   & 33,369 & 100.79 & 125.15 & 120.40 & Table \ref{tab:mseqa} \\
& ProofWiki & Binary  & 1,099 & 15,763 & 1.03 & 48.37 & 196.87   & Table \ref{tab:proofwikiqa}\\
& Stacks    & Binary &  776  & 10,423 & 1.00  & 55.47 & 171.07 & Table \ref{tab:stacksqa} \\
\midrule
\multirow{4}{*}{Premise Retrieval}
& NaturalProofs \cite{welleck2021naturalproofs}     & Binary &  2,060 & 40,806 & 3.94 & 49.51 & 62.32 & Table \ref{tab:naturalproofs} \\
& LeanDojo \cite{yang2023leandojo}        & Binary &  4,109  & 180,944 & 2.33 & 106.28 & 30.18 & Table \ref{tab:leanps} \\
& MAPL \cite{mikula2024magnushammer} & Binary & 4,000 & 493,029 & 7.07 & 43.53 & 30.15  & Table \ref{tab:isabelleps}\\
& HolStep \cite{kaliszyk2017holstep} & Binary & 1,411   & 3,973 & 22.82 & 34.33 & 28.84 & Table \ref{tab:holps} \\
\midrule
\multirow{2}{*}{Formula Retrieval}
& NTCIR-WFB \cite{zanibbi2016ntcir}  & 3-level &  39   & 1,994 & 38.95 & 2.72 & 2.93 & Table \ref{tab:ntcir} \\
& ARQMath-Task-2 \cite{10.1007/978-3-030-58219-7_15,10.1007/978-3-030-85251-1_17,10.1007/978-3-030-99739-7_51}  & 4-level &  76   & 9,969 & 63.18 & 122.25 & 5.61 & Table \ref{tab:mseformula} \\
\bottomrule
\end{tabular}}
\end{table}

\subsection{Semantic Statement Retrieval}\label{subsec:ssr}
Semantic Statement Retrieval is the task of retrieving semantically similar statements or questions given a math query, which itself is a mathematical statement or question. This task is motivated by real-world scenarios such as searching for theorems in mathematical libraries—for example, users of Lean often need to look up theorems in mathlib4. One instance of this task is Informalized Mathlib4 Retrieval, where the goal is to retrieve relevant mathlib4 theorems based on informal mathematical queries. Another instance is Duplicate Question Retrieval, which involves retrieving questions labeled as duplicates on math forums like Mathematics Stack Exchange (MSE) and Math Overflow (MO). This task is inspired by the CQADupStack dataset \cite{10.1145/2838931.2838934}. A key challenge in this task is identifying semantically equivalent questions that may differ in phrasing or notation but express the same mathematical meaning. We construct two datasets for this purpose: MSE Duplicate Question Retrieval and MO Duplicate Question Retrieval. The details of all three datasets are discussed in the following paragraphs.

\paragraph{Informalized Mathlib4 Retrieval.} We use the evaluation dataset from \cite{gao-etal-2024-semantic-search}. The original dataset contains both formal and informal queries; in this work, we focus only on the informal queries, retaining 40 out of the original 50. The retrieval corpus consists of informalized mathlib4 statements. Relevance is graded on a three-level scale, with the criteria defined in the original paper. An example query and its relevant document are shown in Table \ref{tab:inf_mathlib}.

\begin{table}[t]
\caption{Informalized Mathlib4 Retrieval example.}
\label{tab:inf_mathlib}
\centering
\begin{tabular}{p{0.45\linewidth}p{0.45\linewidth}}
\toprule
\textbf{Query} & \textbf{Relevant Document} \\
\midrule
Let $L/K$ be a Galois extension, $F$ be an intermidiate field, then $L^{\{\sigma \in \text{Gal}(L/K) | \sigma x = x , \forall x \in F \} } = F$ & Fixed Field of Fixing Subgroup Theorem: For a Galois field extension E/F with an intermediate field K, the fixed field of the subgroup fixing K is equal to K. \\
\bottomrule
\end{tabular}
\end{table}

\paragraph{MSE Dup. Question Retrieval.} The task of Duplicate Question Retrieval involves retrieving questions that are duplicates of a given input question. We construct our dataset using the Mathematics Stack Exchange Data Dump (2024-09-30)\footnote{\href{https://archive.org/download/stackexchange_20240930/stackexchange_20240930/math.stackexchange.com.7z}{https://archive.org/download/stackexchange\_20240930/stackexchange\_20240930/math.stackexchange.com.7z}}. We begin by extracting all question posts and removing those containing figures, links, or tables. Next, we build an undirected graph where an edge connects two questions if they are marked as duplicates in the data dump. We compute the transitive closure of this graph to ensure that if question A is a duplicate of B and B is a duplicate of C, then A is also considered a duplicate of C. From each connected component in the graph, we randomly sample one question to serve as a query. The remaining questions constitute the initial corpus, which we further refine. To mitigate the issue of false negatives—questions that are duplicates but not labeled as such—we adopt a dynamic corpus approach similar to the LeetCode dataset in BRIGHT \cite{su2025bright}. Specifically, we extract the tags for each question from the data dump. For a query $Q$ with tag set $T(Q)$, we exclude a candidate question $Q'$ from its corpus if the tag overlap satisfies $\frac{|T(Q)\cap T(Q')|}{|T(Q)|} \ge 0.5$. This ensures that, aside from the ground-truth duplicates, most questions in the corpus are not on the same topic as the query, thus reducing the risk of unlabeled duplicates appearing as false negatives.

\begin{table}[t]
\caption{MSE Dup. Question Retrieval example.}
\label{tab:msedup}
\centering
\begin{tabular}{p{0.45\linewidth}p{0.45\linewidth}}
\toprule
\textbf{Query} & \textbf{Relevant Document} \\
\midrule
Example of divisor $D$ such that $\deg D >0$ and $\ell(D)=0$ It is easy to see that if a divisor $D$ on a projective curve $C$ over a field $K$ has negative degree, then $\ell(D) = \dim_K \{f \in K(C) \mid div(f)+D\ge 0\}$ is zero. However, I suppose that the converse is not true. Can someone give me the simplest example of a divisor $D$ on some curve $C$ satisfying $\deg(D)>0$ but $\ell(D)=0$? & Does the dual of a line bundle with no sections have a section? Let $L \to X$ be a holomorphic line bundle over a compact complex manifold. Suppose $L$  is non-trivial and has no non-trivial sections. Let me ask the following (hopefully not entirely trivial) question: Does the dual $L^{\ast}$ have a non-trivial section? A special case of this is when $L$ is the dual of an ample line bundle. Obviously ample line bundles have sections, but the dual does not.\\
\bottomrule
\end{tabular}
\end{table}

\paragraph{MO Dup. Question Retrieval.} The construction of the MO Duplicate Question Retrieval dataset follows the same procedure as for the MSE dataset. We use the MathOverflow Data Dump (2024-09-30)\footnote{\href{https://archive.org/download/stackexchange_20240930/stackexchange_20240930/mathoverflow.net.7z}{https://archive.org/download/stackexchange\_20240930/stackexchange\_20240930/mathoverflow.net.7z}}. After cleaning the question posts, applying transitive closure to the graph, and filtering the corpus, we obtain 225 queries and 108,301 documents.

\subsection{Question-Answer Retrieval}\label{subsec:qar}
Question-Answer Retrieval focuses on retrieving relevant answers or proofs for a given mathematical question. The main challenge lies in understanding the underlying mathematical intent of the question and identifying documents that provide accurate and precise answers—an objective that goes beyond simple semantic similarity. We include three datasets for this task: ARQMath-Task-1, ProofWiki, and Stacks, which are discussed in the following paragraphs.

\paragraph{ARQMath-Task-1.} ARQMath-Task-1 \cite{10.1007/978-3-030-58219-7_15,10.1007/978-3-030-85251-1_17,10.1007/978-3-030-99739-7_51} is an answer retrieval task, where the goal is to retrieve relevant answer posts from Mathematics Stack Exchange (MSE) between 2010 and 2018, given a query question posted after 2019. The task was held over three years, with the query sets consisting of MSE questions from 2019, 2020, and 2021, respectively. We use ARQMath-3-Task-1 as the test set. The ARQMath-3-Task-1 dataset contains 78 queries, with an average of 446.8 annotated answers per query. Relevance is graded on four levels, and readers may refer to \cite{10.1007/978-3-030-99739-7_51} for the detailed relevance criteria. The evaluation metric is nDCG-prime, introduced in \cite{sakai2008information}, which excludes unjudged documents from the ranked list. As a result, we adopt a dynamic corpus approach, where the corpus for each query consists only of its associated annotated documents.

\begin{table}[t]
\caption{ARQMath-Task-1 example.}
\label{tab:mseqa}
\centering
\begin{tabular}{p{0.45\linewidth}p{0.45\linewidth}}
\toprule
\textbf{Query} & \textbf{Relevant Document} \\
\midrule
Confusion about the formula of the area of a surface of revolution Before I read the formula of the area of revolution which is $\int 2\pi y \,ds$, where $ds = \sqrt{1 + \frac{dy}{dx}^2}$, I thought of deriving it myself. I tried to apply the same logic used for calculating the volume of revolution (e.g., $\int \pi y^2 dx $). My idea is to use many tiny hollow cylinders (inspired from the shell method), each has a surface area of $(2\pi y) (dx)$:  $2\pi y$ is the circumference of the cylinder, and $dx$ is the height of the cylinder  Their product is the surface area of the hollow (e.g., empty from the inside) cylinder. With this logic, the area is $\int 2\pi y dx$. Where is my mistake? Also it's confusing why for the volume it was enough to partition the object using cylinders and for areas not. & You should review the formula for the surface area in the case of a surface of revolution (e.g. here). The surface area of the surface obtained by rotation the graph of $y=f(x)$ about the $x$-axis on the interval $[x_1,x_2]$, is given by: $2\pi\int_{x_1}^{x_2} y \sqrt{1+ \left(y'\right)^2}\,\mbox{d}x = 2\pi\int_{x_1}^{x_2} f(x) \sqrt{1+ \left(f'(x)\right)^2}\,\mbox{d}x$ Now if $f(x) = \tfrac{\cosh(4x)}{4}$, then $f'(x)=\sinh(4x)$ so rotation on $[-1,1]$ gives: $\frac{\pi}{2}\int_{-1}^{1} \cosh(4x) \sqrt{1+ \sinh^2(4x)}\,\mbox{d}x$ You can simplify (a lot). Can you take it from here?   I also need to know how would one go about rotating this about the y-axis, but have no idea where to start.  The link from above also covers the formula for rotation about the $y$-axis.\\
\bottomrule
\end{tabular}
\end{table}

\paragraph{ProofWiki.} ProofWiki is a mathematical library containing definitions, axioms, theorems, and their corresponding proofs. In the ProofWiki Question-Answer Retrieval task, the queries are theorems from ProofWiki, and the corpus consists of proofs sourced from the same platform. The objective is to retrieve the correct proof(s) for a given theorem. Since some theorems in ProofWiki have multiple proofs, the average number of relevant documents per query is greater than one. We use the theorems from the test set of the ProofWiki dataset in NaturalProofs \cite{welleck2021naturalproofs} as queries, and include all proofs from the dataset, not just those associated with the queries, as the retrieval corpus.

\paragraph{Stacks.} The Stacks Project is a mathematical library focused on algebraic stacks and algebraic geometry. Similar to the ProofWiki Question-Answer Retrieval task, Stacks Question-Answer Retrieval aims to retrieve the correct proof for a given theorem in the Stacks Project. We use theorems from the test set of the Stacks dataset in NaturalProofs \cite{welleck2021naturalproofs} as queries, and include all proofs from the dataset as the retrieval corpus.

\subsection{Premise Retrieval}\label{subsec:pr}
Premise retrieval is the task of retrieving definitions, theorems, and lemmas that are useful for proving a target theorem or advancing the current proof state. This task plays a crucial role in automated theorem proving, where the ability to efficiently identify relevant mathematical premises can greatly influence the success of the proof process \cite{mikula2024magnushammer,yang2023leandojo}. We include four datasets for this task: one natural language premise retrieval dataset, NaturalProofs \cite{welleck2021naturalproofs}, and three formal premise retrieval datasets: LeanDojo \cite{yang2023leandojo} for Lean, MAPL \cite{mikula2024magnushammer} for Isabelle, and HolStep \cite{kaliszyk2017holstep} for HOL Light. The details of these four datasets are discussed in the following paragraphs.

\paragraph{NaturalProofs.} NaturalProofs \cite{welleck2021naturalproofs} is a natural language premise retrieval dataset, where the goal is to retrieve definitions, lemmas, and theorems that are useful for proving a given query statement. It consists of four subsets: ProofWiki, Stacks, Real Analysis, and Number Theory. In the ProofWiki subset, the query is a theorem from ProofWiki, the corpus includes all definitions, lemmas, and theorems in the library, and the relevant documents are those used in the proof of the query theorem. The other three subsets follow a similar formulation. We evaluate each subset separately and report the average of their scores as the final result for the NaturalProofs dataset.

\begin{table}[t]
\caption{NaturalProofs example.}
\label{tab:naturalproofs}
\centering
\begin{tabular}{p{0.45\linewidth}p{0.45\linewidth}}
\toprule
\textbf{Query} & \textbf{Relevant Document} \\
\midrule
If ${\mathcal H}$ is an open covering of a closed and bounded subset $S$ of the real line$,$ then $S$ has an open covering $\widetilde{\mathcal H}$ consisting of finitely many open sets belonging to ${\mathcal H}.$ & no point of $S^c$ is a limit point of~$S.$ \\
\bottomrule
\end{tabular}
\end{table}

\paragraph{LeanDojo.} LeanDojo \cite{yang2023leandojo} provides a premise retrieval dataset for Lean, where the goal is to retrieve useful premises from mathlib4 to advance a given Lean 4 proof state. In this task, the query is a proof state, the corpus consists of all mathlib4 declarations, and the relevant documents are the premises used in the next tactic step. We follow the \verb+novel_premises+ data split from the original benchmark, in which each proof in the test set uses at least one premise not seen during training.

\paragraph{MAPL.} MAPL \cite{mikula2024magnushammer} is a premise retrieval dataset for Isabelle. The task is similar to that of LeanDojo premise retrieval, where the goal is to retrieve useful premises to advance the current proof state. In MAPL, the query is an Isabelle proof state and the corpus consists of premises expressed in Isabelle's formal language. The original dataset comprises a collection of (state, premise) pairs, which we split into train, dev, and test sets following a strategy similar to the \verb+novel_premises+ split in LeanDojo. Specifically, each proof state in the test set uses at least one premise that does not appear in the training set.

\paragraph{HolStep.} HolStep \cite{kaliszyk2017holstep} is a dataset based on HOL Light proofs. Each file in the original dataset contains a single conjecture along with the dependencies used in its proof. We treat the conjectures as queries and aggregate all dependencies across the dataset to form the retrieval corpus. The task is to retrieve the relevant dependencies for a given conjecture.

\subsection{Formula Retrieval}\label{subsec:fr}
Formula retrieval focuses on retrieving mathematical expressions that are relevant to a given query formula, optionally incorporating the formula’s surrounding context. This task requires a deep understanding of the semantic meaning of mathematical formulas. We evaluate this task using two datasets: NTCIR-12 Wikipedia Formula Browsing (WFB) \cite{zanibbi2016ntcir} and ARQMath-Task 2 \cite{10.1007/978-3-030-58219-7_15,10.1007/978-3-030-85251-1_17,10.1007/978-3-030-99739-7_51}.

\paragraph{NTCIR-WFB.} The NTCIR-12 Wikipedia Formula Browsing task involves retrieving relevant formulas given a query formula. The corpus consists of mathematical formulas extracted from Wikipedia articles. Relevance is graded on a three-level scale, with detailed criteria provided in \cite{zanibbi2016ntcir}. Similar to ARQMath-Task-1, we adopt a dynamic corpus approach, where each query is evaluated against only its associated annotated documents.

\begin{table}[t]
\caption{NTCIR-WFB example.}
\label{tab:ntcir}
\centering
\begin{tabular}{p{0.45\linewidth}p{0.45\linewidth}}
\toprule
\textbf{Query} & \textbf{Relevant Document} \\
\midrule
$L(\lambda,\alpha,s)=\sum_{n=0}^{\infty}\frac{\exp(2\pi i\lambda n)}{(n+\alpha)^{s}}.$& $g(s)=\sum_{n=1}^{\infty}\frac{a(n)}{n^{s}}$ \\
\bottomrule
\end{tabular}
\end{table}

\paragraph{ARQMath-Task-2.} ARQMath-Task-2  \cite{10.1007/978-3-030-58219-7_15,10.1007/978-3-030-85251-1_17,10.1007/978-3-030-99739-7_51} is a formula retrieval task, where the goal is to retrieve relevant formulas from MSE posts given a query formula along with its context (i.e., the question post in which it appears). We use ARQMath-3-Task-2 as the test set, which contains 76 queries and an average of 63.18 annotated relevant documents per query. The task defines four levels of relevance, with criteria detailed in \cite{10.1007/978-3-030-99739-7_51}. Similar to ARQMath-Task-1, we adopt a dynamic corpus approach, where each query’s corpus consists only of its annotated documents.

\section{Experiments}\label{sec:exp}
In this section, we evaluate the performance of 13 retrieval models on MIRB. The experimental setup is described in SubSection \ref{subsec:exp_setup}, and the comparison of model performance is presented in SubSection \ref{subsec:main_results}.

\subsection{Experiment Setup}\label{subsec:exp_setup}
We evaluate four groups of retrieval models. For the sparse model, we test BM25. For open-source models with fewer than 1 billion parameters, we include gte-large-en-v1.5 \cite{li2023towards}, UAE-Large-V1 \cite{li-li-2024-aoe}, and bge-large-en-v1.5 \cite{xiao2024c}. For open-source models with more than 1 billion parameters, we evaluate gte-Qwen2-1.5B-instruct \cite{li2023towards}, e5-mistral-7b-instruct \cite{wang-etal-2024-improving-text}, NV-Embed-v2 \cite{lee2025nvembed}, gte-Qwen2-7B-instruct \cite{li2023towards}, SFR-Embedding-2\_R \cite{SFR-embedding-2}, and GritLM-7B \cite{muennighoff2024generative}. For proprietary models, we evaluate Cohere-embed-english-v3.0\footnote{\href{https://huggingface.co/Cohere/Cohere-embed-english-v3.0}{https://huggingface.co/Cohere/Cohere-embed-english-v3.0}}, text-embedding-3-large \footnote{\href{https://platform.openai.com/docs/models/text-embedding-3-large}{https://platform.openai.com/docs/models/text-embedding-3-large}}, and voyage-3-large\footnote{\href{https://huggingface.co/voyageai/voyage-3-large}{https://huggingface.co/voyageai/voyage-3-large}}.

For dense models, we compute the cosine similarity between the query embedding and the corpus embeddings, and return a ranked list of documents. Model configurations, including the maximum context length for queries and documents, as well as whether instructions are prepended to the queries, are provided in Table~\ref{tab:model_config}. The instructions used are listed in Table~\ref{tab:model_inst}. Following prior work \cite{thakur2021beir,su2025bright}, we report nDCG@10 as the main evaluation metric.

\begin{table}[tb]
\caption{Model configuration. Max $|Q|$ and Max $|D|$ is the maximum context length we set for each model. The instruction column denotes whether we prepend instructions to the query.}
\label{tab:model_config}
\centering
\resizebox{0.5\linewidth}{!}{
\begin{tabular}{lcccc}
\toprule
  & \textbf{Size} & \textbf{Max} $|Q|$  & \textbf{Max} $|D|$ & \textbf{Instruction} \\
\midrule
\multicolumn{5}{c}{\textit{Sparse model}} \\
\midrule
BM25 & - & - & - & No \\
\midrule
\multicolumn{5}{c}{\textit{Open-source models (\textless{}1B)}} \\
\midrule
gte-large-en-v1.5 & 434M & 8192 & 8192 &  No \\
UAE-Large-V1 & 335M & 512 & 512 &  Yes  \\
bge-large-en-v1.5 & 335M & 512 & 512  & Yes   \\
\midrule
\multicolumn{5}{c}{\textit{Open-source models (\textgreater{}1B)}} \\
\midrule
gte-Qwen2-1.5B-instruct &1.78B & 4096 & 4096 &  Yes\\
e5-mistral-7b-instruct &7.11B & 4096&4096 &  Yes\\
NV-Embed-v2 & 7.85B& 32768& 32768& Yes \\
gte-Qwen2-7B-instruct & 7.61B &4096 & 4096& Yes \\
SFR-Embedding-2\_R &7.11B &4096 &4096 & Yes \\
GritLM-7B & 7.24B&4096 &4096 & Yes \\
\midrule
\multicolumn{5}{c}{\textit{Proprietary models}} \\
\midrule
Cohere-embed-english-v3.0 & -& 512 & 512 & No \\
text-embedding-3-large & -& 8192 & 8192 & No \\
voyage-3-large &- & 32000 &  32000 & Yes \\
\bottomrule
\end{tabular}}
\end{table}

\begin{table}[t]
\caption{Instructions used for different datasets are applied to all models that utilize instructions, except for UAE-Large-V1 and bge-large-en-v1.5. For these two models, the instruction used is: "Represent this sentence for searching relevant passages:"}
\label{tab:model_inst}
\centering
\resizebox{\linewidth}{!}{
\begin{tabular}{ll}
\toprule
\textbf{Dataset} & \textbf{Instruction} \\
\midrule
Informalized Mathlib4 Retrieval & Given a mathematical query, retrieve relevant theorems. \\
\midrule
MSE Dup. Question Retrieval & \multirow{2}{*}{Given a math question, retrieve questions that are duplicates of the given one} \\
MO Dup. Question Retrieval \\
\midrule
ARQMath-Task-1 & Given a math problem, retrieve its solution. \\
\midrule
ProofWiki & \multirow{2}{*}{Given a math theorem, retrieve its proof. }\\
Stacks \\
\midrule
NaturalProofs & Given a math theorem, retrieve useful references, such as theorems, lemmas, and definitions, that are useful for proving the given theorem. \\
\midrule
LeanDojo & Given a Lean 4 proof state, retrieve the declarations that are useful for proving it. \\
\midrule
MAPL & Given an Isabelle proof state, retrieve the declarations that are useful for proving it. \\
\midrule
HolStep & Given a HOL conjecture, retrieve the declarations that are useful for proving it. \\
\midrule
NTCIR-WFB & Given a math formula, retrieve relevant formulas. \\
\midrule
ARQMath-Task-2 & Given a math formula and its context, retrieve relevant formulas. \\
\bottomrule
\end{tabular}}
\end{table}

\subsection{Results}\label{subsec:main_results}

\paragraph{Main Results}
The results are shown in Table \ref{tab:full_results}. BM25 underperforms compared to dense retrievers, and there is a clear performance gap between small models (fewer than 1B parameters) and larger models (around 7B). voyage-3-large outperforms all other models, achieving an average nDCG@10 score of 54.54 and ranking first on 7 out of the 12 datasets. Among the evaluated tasks, models generally perform better on semantic retrieval tasks such as Semantic Statement Retrieval and Formula Retrieval, while their performance degrades on reasoning-oriented tasks, especially Premise Retrieval. Unlike Question-Answer Retrieval, where the solution or part of it appears in the document, Premise Retrieval requires identifying relevant mathematical statements such as lemmas or theorems that are not part of the answer but are useful for constructing a proof. For formal premise retrieval datasets like LeanDojo, MAPL, and HolStep, embedding models often struggle because they are not extensively pre-trained on large corpora of formal language data. As a result, they are unfamiliar with the notation and syntax of formal languages, and are even less capable of identifying the underlying logical connections between the query state and potential premises. Consequently, even models that perform well on Question-Answer Retrieval (e.g., voyage-3-large) show poor performance on Premise Retrieval. To improve performance on this task, models need to be trained on premise retrieval datasets across different formal languages.

\begin{table}[t]
\caption{The performance of retrieval models in MIRB. We report nDCG@10 for all datasets. Avg. denotes the average score
across datasets. The best score for each dataset is highlighted in bold.}
\label{tab:full_results}
\centering
\resizebox{\linewidth}{!}{
\begin{tabular}{lccccccccccccc}
\toprule
 &  \multicolumn{3}{c}{\textbf{Semantic Statement Retrieval}}  & \multicolumn{3}{c}{\textbf{Question-Answer Retrieval}} & \multicolumn{4}{c}{\textbf{Premise Retrieval}}   & \multicolumn{2}{c}{\textbf{Formula Retrieval}} & \multirow{2}{*}{\textbf{Avg.}}      \\
\cmidrule(r){2-4}\cmidrule(lr){5-7}\cmidrule(lr){8-11}\cmidrule(l){12-13}
 & Informalized Mathlib4 Retrieval & MSE Dup. Question Retrieval & MO Dup. Question Retrieval   & ARQMath-Task-1  & ProofWiki & Stacks &  NaturalProofs  & LeanDojo  & MAPL & HolStep & NTCIR-WFB & ARQMath-Task-2  \\
\midrule
\multicolumn{14}{c}{\textit{Sparse model}} \\
\midrule
BM25 & 31.49  & 22.85 & 44.01 & 24.83 & 57.35 & 35.49 & 24.14 &6.91  & 15.27  & 25.88  & 66.03 & 32.46 & 32.23 \\
\midrule
\multicolumn{14}{c}{\textit{Open-source models (\textless{}1B)}} \\
\midrule
gte-large-en-v1.5 & 38.05  & 46.76 & 68.04 & 37.78 & 66.49 & 32.26 & 28.42 & 3.73 & 8.78 & 29.15 & 68.83 &59.87  &  40.68 \\
UAE-Large-V1 & 40.43 &  41.11 & 67.44 & 31.66 & 54.81 & 28.17 & 27.85 & 4.64 & 5.59 & 30.17 & 71.92 & 55.50 & 38.27 \\
bge-large-en-v1.5 &41.99 & 41.70 &67.40  & 31.02 & 56.36 & 30.25 & 27.53 &5.45  & 6.84 & 30.51 & 73.76 & 55.22 &39.00 \\
\midrule
\multicolumn{14}{c}{\textit{Open-source models (\textgreater{}1B)}} \\
\midrule
gte-Qwen2-1.5B-instruct & 55.17 & 43.13 & 67.73 & 41.97 & 77.83 & 52.56 & 27.46 & 8.40 & 18.64 & 28.05 & 72.96 & 53.56 & 45.62 \\
e5-mistral-7b-instruct & 57.33 & 51.14 & 71.31  & 46.46 & 77.29 & 39.85 & 32.14 & 10.80 & 15.41 & 30.27 & \textbf{78.48} & 57.93 & 47.37 \\
NV-Embed-v2 & 59.48 & 55.00 & 78.47 & 47.34 & 83.08 & 58.56 & 37.21 & 12.27 & 16.58 & \textbf{32.77} & 73.22 & 70.00 & 52.00\\
gte-Qwen2-7B-instruct &40.38 & 38.40 & 61.77 & 44.74 & 77.02 & 49.35 & 30.08 & 11.53 & 17.46 & 28.16 & 77.52 & 54.68 & 44.26 \\
SFR-Embedding-2\_R & \textbf{60.98} & 58.52 &  81.32 & 51.15 & 85.07 & 54.94 & \textbf{34.67} & 11.83 & 17.07 & 30.76 & 75.69 & 65.48 & 52.29 \\
GritLM-7B & 54.09 & 53.05 & 78.60 & 46.35 & 81.59 & 55.89 & 32.92 & 10.68 & 19.53 & 30.80 & 74.22 & 66.56 & 50.36 \\
\midrule
\multicolumn{14}{c}{\textit{Proprietary models}} \\
\midrule
Cohere-embed-english-v3.0 & 42.00 & 42.96 & 61.00 & 38.05 & 66.00 & 32.33 & 28.99 & 6.96 & 13.95 & 29.72 & 73.27 & 54.51 & 40.81 \\
text-embedding-3-large & 49.38 & 52.35 & 76.74 & 45.79 & 81.95 & 56.14 & 31.33 & 11.34 & \textbf{19.94} & 31.02 & 73.06 & 70.18 & 49.93 \\
voyage-3-large & 57.36 & \textbf{60.33} & \textbf{82.87} & \textbf{52.45} & \textbf{91.69} & \textbf{62.62} & 32.74 & \textbf{13.02} & 17.77 & 32.68 & 76.91 & \textbf{74.00} & \textbf{54.54} \\
\bottomrule
\end{tabular}}
\end{table}

\paragraph{Results of Reranking}
Applying rerankers to retrieval results is generally expected to improve performance. To assess their effectiveness on mathematical retrieval tasks, we evaluate two rerankers: bge-reranker-v2-m3 \cite{chen-etal-2024-m3} and jina-reranker-v2-base-multilingual\footnote{\href{https://huggingface.co/jinaai/jina-reranker-v2-base-multilingual}{https://huggingface.co/jinaai/jina-reranker-v2-base-multilingual}}. Each reranker computes a relevance score for the concatenated query and document pair, and then reranks the top 10 retrieved documents accordingly. We apply them to the top five models in MIRB: voyage-3-large, SFR-Embedding-2\_R, NV-Embed-v2, GritLM-7B and text-embedding-3-large, to assess whether reranking improves performance. The results, shown in Table \ref{tab:rerank}, indicate that reranking generally leads to a decline in performance. In a few cases, slight improvements are observed: for example, jina-reranker-v2-base-multilingual raises the score of voyage-3-large on ARQMath-Task-1 from 52.45 to 53.03, and improves SFR-Embedding-2\_R on NTCIR-WFB from 75.69 to 76.13. These results suggest that rerankers trained on general text retrieval tasks may not transfer effectively to mathematical retrieval.

\begin{table}[t]
\caption{Results of reranking. Each $\cdot / \cdot / \cdot$ represents the score before reranking, after applying the bge-reranker-v2-m3, and after applying the jina-reranker-v2-base-multilingual, respectively.}
\label{tab:rerank}
\centering
\resizebox{\linewidth}{!}{
\begin{tabular}{lccccc}
\toprule
 & NV-Embed-v2 &  SFR-Embedding-2\_R & GritLM-7B &  text-embedding-3-large & voyage-3-large \\
\midrule
Informalized Mathlib4 Retrieval &  59.48 / 55.19 / 56.61  &  60.98 / 55.29 / 57.16& 54.09 / 52.39 / 54.71& 49.38 / 46.92 / 48.23& 57.36 / 55.22 / 56.06\\
MSE Dup. Question Retrieval   & 55.00 / 47.23 / 49.00 & 58.52 / 50.28 / 52.25& 53.05 / 46.58 / 48.22 & 52.35 / 46.06 / 47.45& 60.33 / 52.93 / 54.84\\
MO Dup. Question Retrieval & 78.47 / 64.52 / 70.96 & 81.32 / 66.09 / 72.00& 78.60 / 63.70 / 69.00& 76.74 / 61.77 / 67.95& 82.87 / 66.78 / 73.15  \\
\midrule
 ARQMath-Task-1 & 47.34 / 46.66 / 47.17 & 51.15 / 50.41 / 50.52& 46.35 / 43.45 / 44.36 & 45.79 / 43.30 / 44.09 & 52.45 / 51.41 / 53.03 \\
 ProofWiki  & 83.08 / 67.85 / 73.09& 85.07 / 69.29 / 74.05& 81.59 / 67.26 / 72.23& 81.95 / 67.41 / 72.51 & 91.69 / 70.33 / 75.91\\
 Stacks   &58.56 / 44.67 / 51.68& 54.94 / 41.07 / 49.47& 55.89 / 42.03 / 49.60 & 56.14 / 41.87 / 50.93& 62.62 / 45.76 / 53.75 \\
\midrule
 NaturalProofs  &  37.21 / 33.32 / 32.33 & 34.67 / 31.44 / 30.39& 32.92 / 31.00 / 29.76& 31.33 / 29.22 / 28.23& 32.74 / 30.70 / 29.82 \\
 LeanDojo &  12.27 / 10.59 / 11.37& 11.83 / 10.36 / 11.24 & 10.68 / 9.59 / 10.33 & 11.34 / 9.96 / 10.86& 13.02 / 11.10 / 11.83\\
 MAPL  &  16.58 / 16.56 / 16.81 & 17.07 / 17.10 / 17.27 & 19.53 / 18.73 / 18.99& 19.94 / 18.64 / 19.45& 17.77 / 17.06 / 18.08\\
 HolStep   &     32.77 / 31.94 / 31.01 & 30.76 / 30.37 / 29.46 & 30.80 / 30.44 / 29.35 & 31.02 / 30.23 / 29.33 & 32.68 / 31.44 / 30.68 \\

\midrule
 NTCIR-WFB &     73.22 / 71.67 / 73.84  & 75.69 / 74.42 / 76.13  & 74.22 / 73.04 / 73.83 & 73.06 / 71.21 / 72.27 & 76.91 / 74.69 / 76.21\\
 ARQMath-Task-2  &     70.00 / 69.06 / 67.69  & 65.48 / 64.96 / 64.99 & 66.56 / 65.41 / 64.92  & 70.18 / 69.43 / 67.26 & 74.00 / 72.90 / 71.65\\
\midrule
 Avg. &      52.00 / 46.60 / 48.46 & 52.29 / 46.76 / 48.74 & 50.36 / 45.30 / 47.11 & 49.93 / 44.67 / 46.55 & 54.54 / 48.36 / 50.42 \\
\bottomrule
\end{tabular}}
\end{table}

\section{Conclusion}\label{sec:conclusion}
In this paper, we introduce MIRB, a comprehensive benchmark designed to evaluate the mathematical information retrieval capabilities of retrieval models. MIRB comprises four tasks: Semantic Statement Retrieval, Question-Answer Retrieval, Premise Retrieval, and Formula Retrieval. These tasks span both semantic-based and reasoning-based retrieval settings. We evaluate 13 retrieval models and observe that while their performance on semantic-based retrieval is moderate, they perform poorly on reasoning-based tasks. Additionally, applying cross-encoder rerankers does not lead to performance improvements. We hope that MIRB will facilitate future research in mathematical information retrieval and support the development of more effective retrieval models tailored to mathematics.

\section{Limitations}\label{sec:limitations}
Our work has several limitations:
\begin{itemize}
    \item In the ProofWiki Question-Answer Retrieval dataset, we directly use the proofs from ProofWiki as the corpus. A more challenging setup would involve manually adding hard negatives—proofs that appear similar to the ground truth but are not valid proofs for the query theorem. The Stacks Question-Answer Retrieval dataset faces a similar issue.
    \item In the Premise Retrieval datasets, we cover formal languages such as Lean, Isabelle, and HOL. To improve diversity, we should include more formal systems like Coq. However, to the best of our knowledge, there is currently no readily available premise retrieval evaluation dataset for Coq that can be directly incorporated into our benchmark.
    \item The Premise Retrieval datasets are constructed from successful proofs, meaning the ground-truth premises are indeed useful for advancing the proof. However, this does not imply that other premises are not also helpful, leading to potential false negatives. A more ideal approach would involve exhaustively testing all candidate premises with various tactics to determine their utility for the given proof state.
\end{itemize}

\section{Acknowledgment}

This work is supported in part by National Key R\&D Program of China grant 2024YFA1014000 and the New Cornerstone Investigator Program.

\bibliographystyle{plain}
\bibliography{ref}

\newpage
\appendix

\section{Dataset Examples}
In this section, we present examples of datasets from MIRB that are not included in the main text.

\begin{table}[h]
\caption{MO Dup. Question Retrieval example.}
\label{tab:modup}
\centering
\begin{tabular}{p{0.45\linewidth}p{0.45\linewidth}}
\toprule
\textbf{Query} & \textbf{Relevant Document} \\
\midrule
On finite subsets of set of integers, which lies in its sum-set , whose sum of elements equals $0$ Let $n>1$ be an integer and $S \subseteq \mathbb Z$ be such that $|S|=n$ and  $S \subseteq S+S:=\{a+b : a,b \in S\}$ ; then does there exist $T \subseteq S$ with $1 \le |T| \le n/2$ such that $\sum_{a \in T}=0$ ? & Existence of a zero-sum subset Some time ago I heard this question and tried playing around with it. I've never succeeded to making actual progress. Here it goes: Given a finite (nonempty) set of real numbers, $S=\{a_1,a_2,\dots, a_n\}$, with the property that for each $i$ there exist $j,k$ (not necessarily distinct) so that $a_i=a_j+a_k$ (i.e. every element in $S$ can be written as a sum of two elements in $S$, note that this condition is trivially satisfied if $0\in S$ as then every $x\in S$ can be written as $x+0$). Must there exist $\{i_1,i_2,\dots, i_m\}$ (distinct) so that $a_{i_1}+a_{i_2}+\cdots +a_{i_m}=0$? ETA: A possible reformulation can be made in terms of graphs. We can take the vertex set $\{1,\dots ,n\}$ and for each equation $a_i=a_j+a_k$ in S add an edge $[ij]$ and its "dual" $[ik]$. The idea is to find a cycle in this graph, whose dual is a matching. \\
\bottomrule
\end{tabular}
\end{table}

\begin{table}[h]
\caption{ProofWiki example.}
\label{tab:proofwikiqa}
\centering
\begin{tabular}{p{0.45\linewidth}p{0.45\linewidth}}
\toprule
\textbf{Query} & \textbf{Relevant Document} \\
\midrule
Fortissimo Space is not Weakly Countably Compact Let \$T = $\backslash$struct {S, $\backslash$tau\_p}\$ be a Fortissimo space. Then \$T\$ is not weakly countably compact. & 
It suffices to show that $T$ has an infinite subset without limit points. Consider the set $S \setminus \set p$. Let $x \in S$. We have: \{\{begin-eqn\}\} \{\{eqn | l = $\backslash$paren \{S  $\backslash$setminus $\backslash$paren \{S $\backslash$setminus  $\backslash$set p\} \} $\backslash$cup $\backslash$set x       | r = $\backslash$set p $\backslash$cup $\backslash$set x       | c =  \}\} \{\{eqn | r = $\backslash$set \{p, x\}       | c =  \}\} \{\{end-eqn\}\} By definition, $x$ is a limit point of $S \setminus \set p$ {{iff}} $\set {p, x}$ is not a neighborhood of $x$. By definition of Fortissimo space, $\set {p, x}$ is open in $T$. Hence it is a open neighborhood of $x$. Therefore $x$ is not a limit point of $S \setminus \set p$. Since $x$ is arbitrary, $S \setminus \set p$ has no limit points. Hence $T$ is not weakly countably compact. \{\{qed\}\}
\\
\bottomrule
\end{tabular}
\end{table}

\begin{table}[h]
\caption{Stacks example.}
\label{tab:stacksqa}
\centering
\begin{tabular}{p{0.45\linewidth}p{0.45\linewidth}}
\toprule
\textbf{Query} & \textbf{Relevant Document} \\
\midrule
spaces-morphisms-lemma-birational Let $S$ be a scheme. Let $X$ and $Y$ be algebraic space over $S$ with $|X|$ and $|Y|$ irreducible. Then $X$ and $Y$ are birational if and only if there are nonempty open subspaces $U \subset X$ and $V \subset Y$ which are isomorphic as algebraic spaces over $S$. & Assume $X$ and $Y$ are birational. Let $f : U \to Y$ and $g : V \to X$ define inverse dominant rational maps from $X$ to $Y$ and from $Y$ to $X$. After shrinking $U$ we may assume $f : U \to Y$ factors through $V$. As $g \circ f$ is the identity as a dominant rational map, we see that the composition $U \to V \to X$ is the identity on a dense open of $U$. Thus after replacing $U$ by a smaller open we may assume that $U \to V \to X$ is the inclusion of $U$ into $X$. By symmetry we find there exists an open subspace $V' \subset V$ such that $g|_{V'} : V' \to X$ factors through $U \subset X$ and such that $V' \to U \to Y$ is the identity. The inverse image of $|V'|$ by $|U| \to |V|$ is an open of $|U|$ and hence equal to $|U'|$ for some open subspace $U' \subset U$, see Properties of Spaces, Lemma $\backslash$ref\{spaces-properties-lemma-open-subspaces\}. Then $U' \subset U \to V$ factors as $U' \to V'$. Similarly $V' \to U$ factors as $V' \to U'$. The reader finds that $U' \to V'$ and $V' \to U'$ are mutually inverse morphisms of algebraic spaces over $S$ and the proof is complete. \\
\bottomrule
\end{tabular}
\end{table}

\begin{table}[h]
\caption{LeanDojo example.}
\label{tab:leanps}
\centering
\begin{tabular}{p{0.45\linewidth}p{0.45\linewidth}}
\toprule
\textbf{Query} & \textbf{Relevant Document} \\
\midrule
\makecell[tl]{
R : Type u\\
M : Type v\\
inst\textdagger² : CommRing R\\
inst\textdagger¹ : AddCommGroup M\\
inst\textdagger : Module R M\\
B : BilinForm R M\\
f g : Module.End R M\\
hf : IsSkewAdjoint B f\\
hg : IsSkewAdjoint B g\\
$\vdash$ IsAdjointPair B B (f * g) (g * f)} & theorem neg\_mul\_neg (a b : $\alpha$) : -a * -b = a * b\\
\bottomrule
\end{tabular}
\end{table}

\begin{table}[h]
\caption{MAPL example.}
\label{tab:isabelleps}
\centering
\begin{tabular}{p{0.45\linewidth}p{0.45\linewidth}}
\toprule
\textbf{Query} & \textbf{Relevant Document} \\
\midrule
\makecell[tl]{
proof (prove)\\
using this:\\
length ps = length vs\\
left\_nesting f $\backslash\backslash$<noteq> left\_nesting g\\
is\_const (fst (strip\_comb f))\\
goal (1 subgoal):\\
1. match (list\_comb f ps) (list\_comb g vs)\\
= None} 
& 
list\_induct2: fixes xs :: "'c list"   and ys :: "'d list"   and P :: "'c list <Rightarrow> 'd list <Rightarrow> bool" assumes "length xs = length ys"   and "P [] []"   and " <And>x xs y ys. <lbrakk>length xs = length ys; P xs ys <rbrakk> <Longrightarrow> P (x \# xs) (y \# ys)" shows "P xs ys" \\
\bottomrule
\end{tabular}
\end{table}

\begin{table}[h]
\caption{HolStep example.}
\label{tab:holps}
\centering
\begin{tabular}{p{0.45\linewidth}p{0.45\linewidth}}
\toprule
\textbf{Query} & \textbf{Relevant Document} \\
\midrule
ABSOLUTELY\_INTEGRABLE\_CONVOLU\\TION\_LINF\_L1 |- (!bop. (!f. (!g. (!x. (((bilinear bop) $/\backslash$ (((measurable\_on f) UNIV) $/\backslash$ ((bounded ((IMAGE f) UNIV)) $/\backslash$ ((absolutely\_integrable\_on g) UNIV)))) ==> ((absolutely\_integrable\_on ($\backslash$y. ((bop (f ((vector\_sub x) y))) (g y)))) UNIV)))))) & BILINEAR\_SWAP |- (!op. ((bilinear ($\backslash$x. ($\backslash$y. ((op y) x)))) = (bilinear op))) \\
\bottomrule
\end{tabular}
\end{table}

\begin{table}[h]
\caption{ARQMath-Task-2 example.}
\label{tab:mseformula}
\centering
\begin{tabular}{p{0.45\linewidth}p{0.45\linewidth}}
\toprule
\textbf{Query} & \textbf{Relevant Document} \\
\midrule
Formula: $\int \frac{1}{\left(x^2+1\right)^n}dx$ Context: $\int \frac{1}{\left(x^2+1\right)^n}dx$ Let be $n\in \mathbb{Z_+}$. Compute the following integral: $$\int \frac{1}{\left(x^2+1\right)^n}dx$$ I obtained that for $$n=1$$ the value of the integral is $$\tan^{-1}x+C$$ and for $$n=2$$ $$x\left(\frac{1}{2\left(x^2+1\right)}+\frac{\tan \:^{-1}}{2x}\right)+C$$ How to do the rest of the cases?   & $I_n=\int \frac 1 {(x^2-1) ^n}  d x$ \\
\bottomrule
\end{tabular}
\end{table}

\section{Computing Resources}\label{resources}
We conduct our experiments on eight NVIDIA A800 (80G) GPUs. For the sparse model BM25, evaluation on our benchmark takes approximately one hour. For small models with fewer than one billion parameters, evaluation requires around six GPU hours. The 1.5B model takes about 36 GPU hours, while the 7B models require about 64 GPU hours. Each proprietary model is evaluated in under 25 hours.

\section{Broader Impacts}\label{impacts}
We introduce a unified benchmark for mathematical information retrieval, aiming to encourage the development of more effective retrieval models. We hope this benchmark helps advance search engines and automated theorem proving systems by driving progress in math-specific retrieval capabilities.

\end{document}